# Thermal and stress\strain analysis of the tested ITER-like W Langmuir probes in EAST

Chunyu He[a,b], Dahuan Zhu[a,*]

[a] *Institute of Plasma Physics, HFIPS, Chinese Academy of Sciences, Hefei 230031, China*
[b] *University of Science and Technology of China, Hefei 230026, China*

## ABSTRACT

ITER-like tungsten Langmuir probes (W DLPs) have been installed and tested at the lower divertor horizontal target composed of flat-type components in EAST. Due to the non-active cooling, transient thermal and stress\strain analyses considering actual thermal loading and cooling conditions were thus conducted to evaluate the thermal performance and mechanical quality of W DLPs subjected to the long pulse & high plasma flux of EAST. The thermal analysis reveals that the inevitable leading edge induced thermal loading at surrounding area of W DLPs is not ignorable. The thermal performance of W DLPs are largely related to the plasma scenario ($Q_p$: parallel heat flux along magnetic field line, α: incline angle of magnetic field line). Under current plasma parameters, melting of W was not occurred in general, but recrystallization as well as the induced cracks may be still possible. And, the interval period (~1000 s) between neighboring shots is sufficient for nature cooling of W DLPs. The stress analysis also tells that the ceramic LLL may be general a crucial weak point of W DLPs, which is expected to not only limit the thermal affordability of long pulse but also cause possible crack problems.

Such calculation results can provide important reference for current plasma operation and future improvement of the W DLPs.

# 1 Introduction

The divertor is a crucial component in modern tokamak devices, responsible for exhausting plasma heat and low-energy helium ash. To assess and prevent damage to the divertor, divertor Langmuir probes (DLPs) are typically installed on the target plates to measure the plasma parameters directly. All current main tokamaks, including EAST, WEST, ASDEX-Upgrade, DIII-D, JET, JT-60, KSTAR, and others [1-7], are equipped with DLPs, generally using conventional graphite (CFC) as the material for DLPs. However, CFC will sublimate when the surface temperature exceeds 3500 °C. Since 2013, ITER has decided to use a full-tungsten divertor [8], and combined with the situation of the full-metal wall tokamak device like ASDEX-Upgrade and WEST [1, 9], metal DLPs are now becoming more interesting. The ITER Organization has considered tungsten probes with good heat transfer ability (melting point: ~3400 °C, heat transfer coefficient: ~170 $Wm^{-1}K^{-1}$) [10], and recently tantalum probes have been tested in WEST [1]. Additionally, prototypes of tungsten DLPs have been manufactured and passed through 500 cycles of a 15 $MW/m^2$ electron-beam high heat flux test using EMS-60 [11].

To test the thermal and mechanical properties in a tokamak conditions, such ITER-like tungsten probes were installed in EAST for the first time, marking the first instance of their installation in a tokamak device. EAST can provide a plasma current of 300-600 kA with a steady parallel heat flux that varies from 40-120 MW/m$^2$, making it a suitable platform to test the performance of such W DLPs that ITER may use in the future. It is important to emphasize that practical discharge events in a tokamak can provide an evaluation of the impact of lateral high-power plasma heat flux along magnetic field lines, often reaching tens even hundreds of MW/m$^2$, resulting from the inlet angle and probe structure design. This is a capability that is often not available in a normal electron high heat flux test platform.

Combined EAST's recent long pulse and high-power capacities and objectives, there are high risk of damages including melting and cracking under the no active cooling conditions. So, assessment of the damage threshold conditions which was combined with the actual operation conditions is one of the main focuses of this work. And considering ITER's more crucial heat condition, the data here can provide reference of performance boundary of W DLPs.

To provide a clearer understanding of the thermal and mechanical performance of W DLPs under the plasma heat flux exposure in EAST, the following paper is structured into four sections. Section 2 covers the detail structure and materials of the W DLPs in EAST, while Section 3 explains the ANSYS analysis settings. Section 4 presents the results and corresponding discussions. And section 5 is the summaries and conclusions.

## 2 Tested W DLP installed in lower divertor in EAST

2.1 Structure and materials of tested W DLP

The W DLPs used in EAST have a specific structure, which is depicted in **Figure 1**, The main properties of W and ceramic at different temperatures are listed in **Table 1** and **Table 2**, respectively.

**Figure 1**. The structure of W DLP. (a) one typical W DLP installed in EAST; (b) the 1/2 structure model showing clear cross-section view.

2.2 W DLPs installed in lower divertor

**Figure 2** displays the W DLPs' appearance before the discharge campaign. The W DLPs are placed between two flat-type components (cassettes), which have a semi-circular design to contain them. Using a 3D profile measuring instrument, the maximum protrusion of the W DLPs was found to be about towards the lowest edge of the cassette, which is required for subsequent ANSYS analysis. **Figure 3** shows the installation section of the W DLPs, where the tungsten SSS is isolated from the flat-tile cassette. and thus, during the seconds long heat loading process, it is assumed to be without any cooling conditions. From **Figure 3**, it can be observed that the tungsten SSS's inner and outer lateral surfaces will be subjected to the plasma impact, taking the shape of an ellipse (or almost ellipse).

**Figure 2**. W DLPs installed in EAST

**Figure 3**. Installation section of W DLPs in EAST.

# 3 Modeling

Due to the considerable difference in thermal conductivity coefficients between 92% α-$Al_2O_3$ ceramic and pure tungsten. Consequently, the tungsten element primarily dissipates heat through radiation. Therefore, the ANSYS analysis model shown in **Figure 1(b)** is considered reasonable, even though the heat transfer of the copper component is neglected, as the probe's melting point, particularly that of the tungsten element, is the main concern. As sufficient material data for

92% α-Al$_2$O$_3$ ceramic are not available, the material parameters for 90% α-Al$_2$O$_3$ ceramic presented in **Table 1** are used to define the model's. Although the former has a higher heat transfer coefficient, both materials have much lower values than that of tungsten and, thus, can be assumed to provide heat isolation. Consequently, using 90% α-Al$_2$O$_3$ ceramic instead of 92% α-Al$_2$O$_3$ ceramic will result in a slightly temperature difference, which is an acceptable compromise.

**Figure 3** illustrates that only a specific range of inlet angles can generate sufficient heat flux to reach the top surface of the W DLP. Taking into account the dimensions and placement of the flat-type components and DLP, the heat flux that can reach the pin needs to have an inlet angle α ≥ 2.49°; this means when the plasma current platform $I_p$ ≤ 400 kA (α=2.2°), the W DLP will only suffer the plasma radiation power (about 0.05 MW/m$^2$ in EAST) thus survive all the discharge duration. Based on the operational conditions of the EAST device, **Table 3** lists the parallel heat flux $Q_p$ and equivalent vertical heat flux $Q_v$ values applied to the top surface of the W DLP for various inlet (tilt/incident) angles. Here, α=2.8° and α=3.1° corresponds to $I_p$= 500kA and 600 kA, respectively.

**Table 3.** Heat flux $Q_v$ on the top surface of the W DLP

| $Q_p$(MW/m$^2$)\Inlet angle α | 2.8° | 3.1° |
|---|---|---|
| 40 | 1.95 | 2.16 |
| 60  | 2.93 | 3.24  |
| 70  | 3.42 | 3.78  |
| 80  | 3.91 | 4.32  |
| 100 | 4.88 | 5.41  |
| 120 | 5.86 | 6.49  |
| 150 | 7.32 | 8.11  |
| 200 | 9.77 | 10.81 |

**Figure 4(a)** illustrates the heat loading distribution on the DLP models at an inlet angle of 3.1°. Each pulse includes a 1-second ramp-up period followed by a constant heat flux. If a pulse off condition is present, the pulse falls down instantly. **Table 4** presents the equivalent heat flux for the case where α= 3.1° and $Q_p$=100 MW/m². In ITER, careful consideration should be given to the plasma radiation power, which can reach several MW/m² [12]. However, during a normal discharge event in EAST, the typical plasma radiation power is only about 0.05 MW/m², far smaller than that of ITER.

Unless otherwise specified, the W DLP is assumed to undergo a no-cooling condition during the loading process in all cases discussed in this paper. $Q_v$ is calculated using Function 1:

$$Q_v = Q_p \sin(\alpha) + Q_R \qquad (1)$$

Plasma radiation power: $Q_R$=0.05 MW/m² or 0 MW/m². The heat flux on the lateral surface varies depending on its location. To simplify the analysis, the equivalent lateral heat flux $Q_l$ for the areas $Q_A$ to $Q_D$ is calculated using Function 2:

$$Q_l = Q_p \cos(\alpha) \frac{S_v}{S_l} + Q_R \qquad (2)$$

$S_l$ represents the lateral surface area, while $S_v$ is the projection of $S_l$ onto a plane that is normal to the vector $\mathbf{Q_p} \cos\alpha$.

The effect of mesh was analyzed, and after a mesh sweep, the refined mesh settings are shown in **Figure 4(b)** and **Figure 4(c)**. The mesh on the lateral surface exposed to heat flux was further refined.

**Figure 5** depicts the cooling and displacement boundaries. The surface radiation emissivity of the W DLP is highly dependent on its surface condition. However, during EAST discharge events, the outer surfaces of the W DLP undergoes wall-lithium processing, impurity deposition, and transposition, which can roughen the surface and increase the emissivity coefficient. Regardless of other factors, a conservative emissivity value of 0.3 was chosen based on calibration data from the infrared camera. The outer surfaces (the red surfaces of **Figure 5(a)**) of the W DLP are assumed to radiate at 22°C ambient temperature, while the two cycle surfaces (the blue surfaces of **Figure 5(a)**) are set to 22°C, assuming rapid cooling of the steel screw to room temperature. The pin's displacement in the z direction and the circle surfaces'(red surface of **Figure 5(b)**) displacement in the x direction are set to 0 mm. Moreover, the lateral surfaces (green surfaces of **Figure 5(b)**) of the cylinder will not move in the normal direction.

**Figure 4.** Heat loading surfaces and mesh setting. (a) heat loading on surfaces for case α=3.1°; (b) mesh for analysis; (c) refine mesh on the top of W DLP

**Table 4.** Equivalent heat flux at different surface, α= 3. 1°, $Q_p$=100 MW/m²

| Note | Equivalent Heat flux (MW/m²) |
|:---:|:---:|
| $Q_V$ | 5.41+0.05 |
| $Q_A$ | 52.30+0.05 |
| $Q_B$ | 86.54 |
| $Q_C$ | 45.39 |
| $Q_D$ | 50.58 |
| $Q_E$ | 0.05 |

**Figure 5**. The cooling and displacement boundaries.

# 4. Result and discussion

4.1 Lateral heat flux's influence on temperature distribution

**Figure 6** depicts the impact of lateral heat flux on the temperature distribution when the maximum temperature reaches the melting point (3400°C) of material W under discharge conditions characterized by $\alpha$ = 3.1° and $Q_p$ = 100 MW/m². The orientation of the equal temperature surfaces is significantly altered by the lateral heat flux, which in turn strongly influences the thermal stress. The lateral heat flux is caused by the complex changes in the surface area of the W DLP facing the plasma due to the 3D nature of the equal temperature surfaces. As a result, the calibration of the W DLP's surface area is physically affected by the lateral heat flux.

It is worth emphasizing that the clearance between the shell and pin can makes great sense to the lateral heat flux influence. For case $\alpha$=2.8° and $Q_p$=100 MW/m², the outer lateral heat flux of the shell can be neglected due to its negligible 0.00 mm² projection area. Thus, only the inner lateral heat flux is considered. Analysis results shows that the inner lateral heat flux shortens the melting time threshold from 202.65s (no lateral heat flux condition) to 160.23s.

In summary, it means that the W DLP's design should consider the lateral heat flux effect carefully, even if the outer shell lateral heat flux is only 0.1mm long, it can make a significant difference in the melting time threshold, and only 0.5mm clearance between the pin and shell makes sense to the lateral heat flux effect. Finally, although it was announced in 2022 that a proposal for ITER W DLPs had successfully passed a 15 MW/m$^2$ steady electron high heat flux test for 500 cycles [11], there is still a significant amount of work that needs to be evaluated to consider the lateral heat flux problems in actual tokamak devices.

**Figure 6**. temperature distributions by XZ plane of case α=3.1°, $Q_p$=100MW/m$^2$ at 36.02s; (a) with lateral heat flux ;(b) without lateral heat flux.

4.2 melting time threshold and stress analysis.

As **Figure 6(a)** shows, the maximum temperature of the shell of the W DLP is higher than that of the pin, while the ceramic LLL is hotter than the table. Hence, only the melting time threshold of these components needs to be considered. EAST has exhibited a consistent average discharge power of 40-120 MW/m$^2$, with occasional power fluctuations measured up to 160 MW/m$^2$ by carbon DLP. Therefore, an analysis of a range of parallel powers from 40-200 MW/m$^2$ is warranted.

In **Figure 7**, the melting time thresholds for tungsten and ceramic materials exposed to discharges of 500 kA and 600 kA are plotted. The solid symbols represent the current long-pulse stable power levels, while the open symbols represent fluctuating or anticipated future power loads. The melting time threshold for the current discharge power level is greater than 22.38 seconds (120MW/m² at 3.1º), which is significantly longer than the current 8-second 500-600 kA discharge setting. Consequently, the W DLP is not anticipated to melt during the current discharge event.

Based on the analysis, tungsten is not expected to melt until $Q_p$ reaches 200 MW/m² under 600 kA discharge conditions. As $Q_p$ increases, the melting time threshold difference between tungsten and ceramic becomes smaller. At an angle of α=3.1º and $Q_p$=150 MW/m², their melting time threshold lines intersect, indicating that tungsten will melt before the ceramic LLL. This observation highlights the existence of a maximum $Q_p$ value beyond which tungsten will melt due to insufficient heat conduction. According to Fourier's Law[13], the linear heat flux $\boldsymbol{q} = \lambda \boldsymbol{grad T}$, assuming $\lambda = 98.7\ W/mK$, $\boldsymbol{grad T} = \frac{3400K}{1cm}$, the maximum heat flux that the W shell can withstand is $q_{max}$=33.56 MW in 1D model, thus $q_{max}$=33.56 MW/m² in 3D model. Since the top surface's heat flux $Q_v \leqslant 0.32 q_{max}$, tungsten will not melt before the ceramic LLL if lateral heat flux is not considered. In conclusion, the intersection point is due to the limited heat conductive coefficient and high lateral heat flux.

**Figure 8** shows the maximum temperature reached after an 8-second full power discharge with a 1-second ramp time, and the RCT (recrystallization) time threshold under this discharge mode for both present and anticipated parallel power. The maximum temperature increases linearly with parallel power for both 500 kA and 600 kA discharges. By linearly fitting and considering tungsten's RCT at around 1250 °C [14], the maximum parallel power that the W DLP can withstand without suffering recrystallization is 156.76 MW/m$^2$ and 74.13 MW/m$^2$ for 500 kA and 600 kA discharges, respectively. Therefore, it can be concluded that the present 500 kA discharge in EAST does not pose a risk of tungsten recrystallization, whereas, for the same purpose, the 600kA discharge should either limit the maximum parallel power to 74.13 MW/m$^2$ or restrict the maximum discharge time to 2.85 seconds.

To ensure safe operation, it is essential to evaluate whether the W DLP can adequately cool down during the discharge interval, which typically lasts from 10 minutes to 1 hour. **Figure 9** demonstrates that even for the most severe present discharge case with α= 3.1º and $Q_p$=120MW/m$^2$, 10 minutes are adequate for the W DLP to cool down to its original temperature, and this is also valid for other cases. Moreover, even for the anticipated 600 kA 200 MW/m$^2$ discharge, where the W DLP cools down from its melting point of 3400ºC to its original temperature of

22°C, a cooling period of 10 minutes is still sufficient. Therefore, the current interval (10 minutes to 1 hour) between neighboring shots is adequate for the natural cooling of W DLPs, thus no accumulate heat effect on such W DLPs in EAST.

**Figure 10** illustrates the evolution of maximum principal stress for the brittle ceramic and the equivalent Von-Mises strain evolution for tungsten under thermal loading conditions plotted in **Figure 9** of $Q_p$=120MW/m$^2$, α= 3.1°. The tensile strength of the ceramic is 267 MPa from 20°C to 500 °C [15], and regard the W as yield when strain is greater than 0.2% (by referring to the 0.2% offset setting of $S_y$ in The Structural Design Criteria for ITER In-vessel Components (SDC-IC) [16]). The figure suggests that the ceramic layer experiences cracking due to heat loading, slightly before the plastic deformation of the tungsten SSS. The table indicates that the crack in the ceramic TTT is followed by the severe plastic deformation of the pin. The shell experiences maximum strain near the end of the heating process, while all other parts reach their maximum stress or strain values during the cooling process. The pin, on the other hand, exhibits two extreme stress values influenced by both heating and cooling processes.

**Figure 11** presents the distributions of maximum principal stress in both the ceramic LLL and the ceramic TTT at the instant of maximum value for case $Q_p$=120MW/m$^2$, α= 3.1°. The red area delineates the zone where the maximum principal stress in the ceramic material exceeds 267 MPa, indicating a risk of imminent failure according to the maximum tensile stress theory. As a result, the upper surface of the ceramic LLL is expected to fracture, while multiple cracks may emerge on the ceramic TTT. It is important to note that the cracks may propagate towards the screw holes, resulting in their enlargement, thereby increasing the risk of mechanical instability during discharge and causing possible vibrations.

**Figure 12** depicts the distribution of equivalent Von-Mises strain at the maximum moment of the shell and the pin. The red area corresponds to the region where tungsten undergoes high risk of plastic deformation. Accordingly, the plastic deformation of the shell can be disregarded. However, the pin will experience severe plastic deformation at the lower boundary between the pin and the table. Ultimately, the junction port of W and ceramic exhibits the highest maximum strain.

**Figure 13** displays a curve that is similar to Figure 11, but for the case of $Q_p$=120MW/m², α= 2.8°. The figure demonstrates that after an 8-second, 500 kA discharge, only the ceramic TTT exhibits cracks. In contrast to a 600kA discharge, for a 500kA discharge, all the maximum values arise during the cooling process. When comparing **Figure 14** to **Figure 11(b)**, it is apparent that only the screw holes are at risk of breaking for the case of $Q_p$=120MW/m², α= 2.8°. All the analysis results above show a fact that lower $I_p$ is good for W DLP's life.

By combining the results of the heat analysis and stress/strain analysis, it becomes clear that the life of the W DPL is mainly limited by the poor heat and mechanical properties of the ceramic TTT and the ceramic LLL. This is because the ceramic part tends to break down well before the tungsten part in present discharge power.

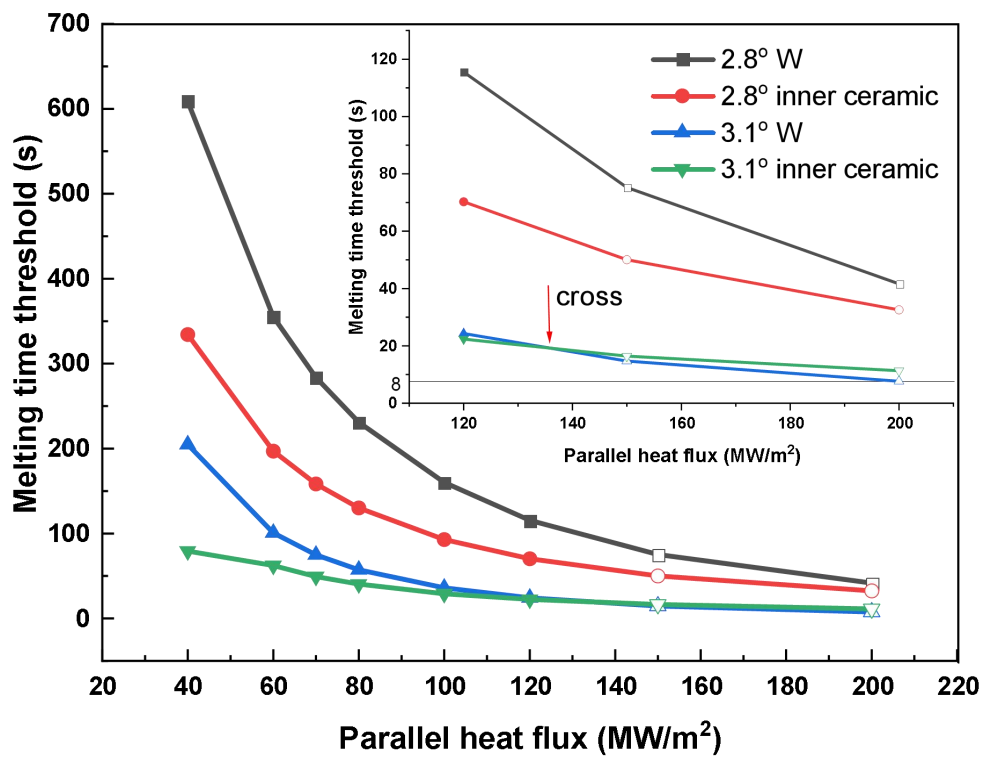

**Figure 7.** W DLP's melting time threshold under different parallel heat flux; solid symbols for present discharge power and open symbols for future.

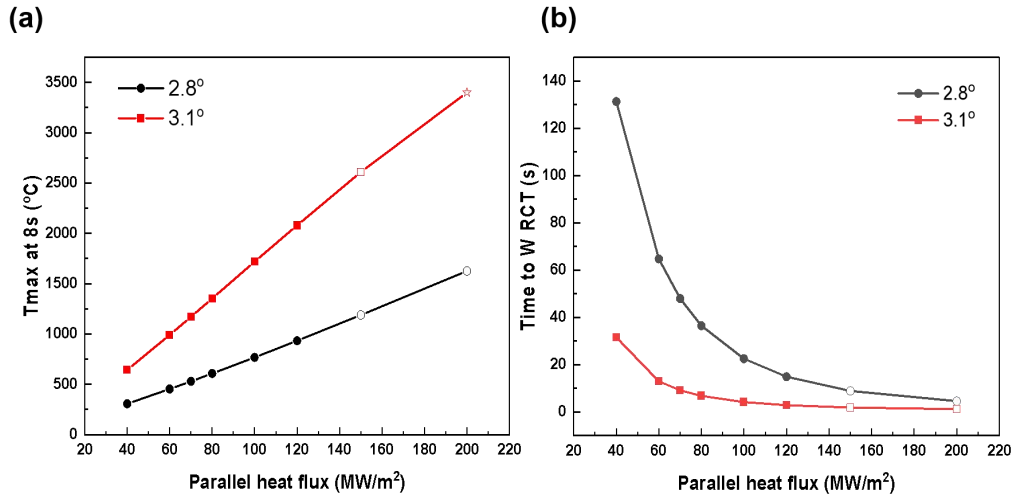

**Figure 8.** (a) W DLP's maximum temperature at 8s and (b) W RCT time threshold; solid symbols for present discharge power and open symbols for future, the star in (a) means the W reach its melting point at 7.63s.

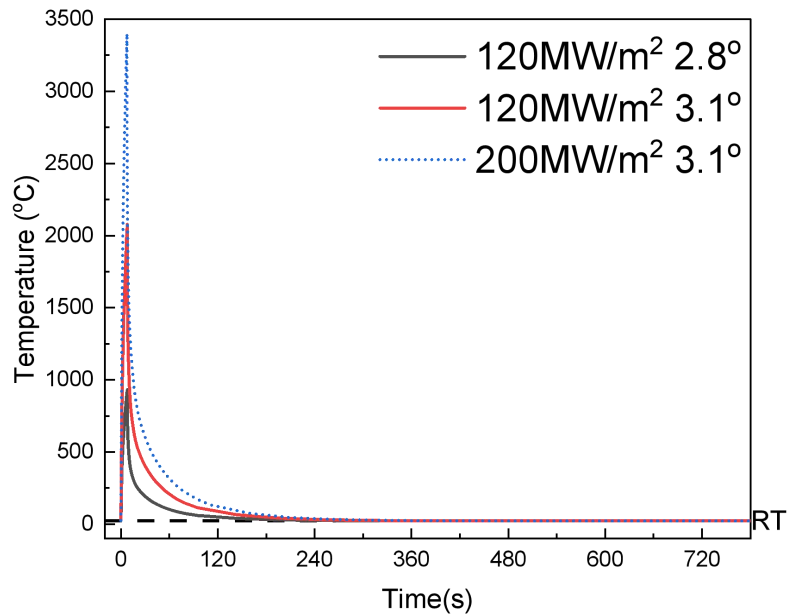

**Figure 9.** W DLP's maximum temperature evolution during 8s heating and 800s cooling process for case $Q_p$=120MW/m$^2$,α= 2.8°;$Q_p$=120MW/m$^2$,α= 3.1° and $Q_p$=200MW/m$^2$,α= 3.1°, solid line for present discharge power and dashed line for anticipate discharge power,

RT=22°C.

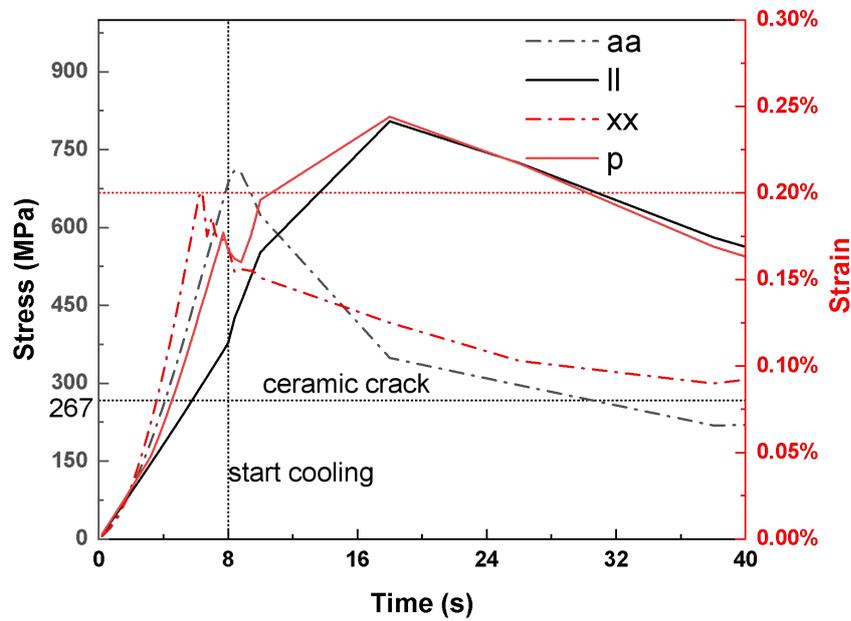

**Figure 10.** Evolution of maximum principal stress (MP) or equivalent Von-Mises strain (VM) for W DLP's different parts at case $Q_p$=120MW/m², α= 3.1°.

**Figure 11.** Maximum principal stress distribution by XZ plane for case α= 3.1° and $Q_p$=120MW/m² at the moment of reaching maximum value, according to **Figure 9**'s stress evolution; (a) ceramic LLL, (b) ceramic TTT, ceramic's tensile strength equals 267MPa.

**Figure 12.** Maximum equivalent Von-Mises strain distribution in same legend for case α= 3.1° and $Q_p$=120MW/m² according to **Figure 9**'s

temperature evolution; (a) section of shell by XZ plane at 6.4s;(b) global model;(c) part of pin by YZ plane at 18s.

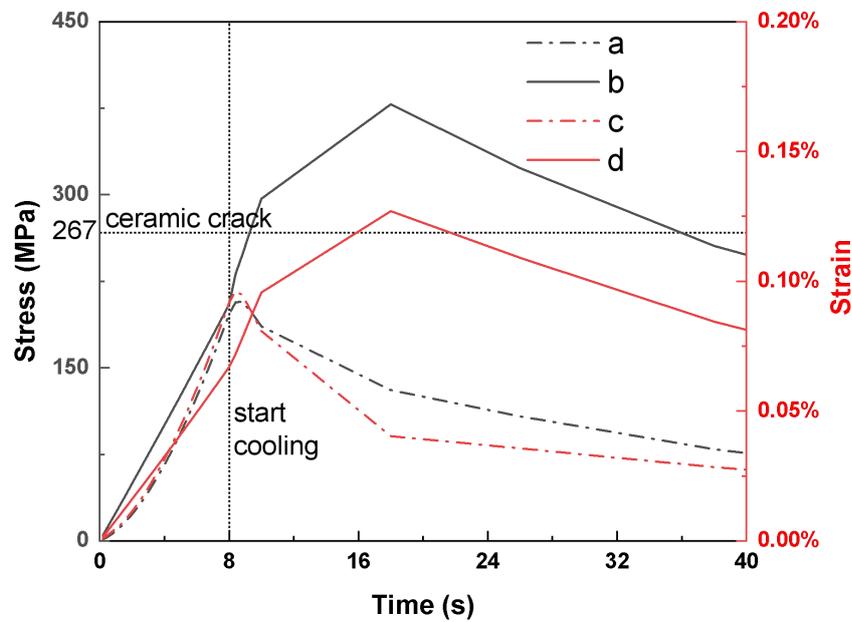

**Figure 13.** Evolution of maximum principal stress (MP) or equivalent Von-Mises strain (VM) for W DLP's different parts at case $Q_p$=120MW/m², α= 2.8º.

**Figure 14.** Maximum principal stress distribution of table by XZ plane at 18s of case $Q_p$=120MW/m², α= 2.8º.

# 5 Conclusion

The thermal analysis of an ITER-like W DLP installed in EAST for the 2022 winter discharge campaign has been conducted. To ensure that the W DLP's pin surface is adequately loaded by plasma, the inlet angle must be greater than 2.49° when the W DLP's head protrudes X mm from the lowest edge of the divertor surface. For EAST's possible inlet angle, the longest tungsten melting time threshold is 608.79 s for case 40 MW/m$^2$ at α=2.8°, while the shortest is 24.29 s for a case 120 MW/m$^2$ at α= 3.1°. The unavoidable lateral inlet heat flux, resulting from the probe's structural design, surprisingly shortens the melting time threshold. Fortunately, considering EAST 500-600KA pulse duration is less than 8s, the W DLPs will not suffer melting problem.

The W DLP will not undergo recrystallization at discharges below 500 kA. However, for discharges at 600 kA, it is necessary to limit the parallel heat flux to 74.13 MW/m$^2$ or the pulse duration to a maximum of 2.85 s (at 120 MW/m$^2$) to avoid the recrystallization problem. For both present discharges at 120 MW/m$^2$ and future discharges up to 200 MW/m$^2$, a cooling time of 10 minutes is sufficient for the W DLP to cool down.

The stress analysis indicates that the ceramic LLL will remain intact at a discharge of 500 kA, while small cracks will develop in the screw holes of the ceramic TTT. However, at a discharge of 600 kA, both the ceramic LLL and the ceramic TTT are susceptible to severe cracking, which warrants significant attention.

Combining the results of the thermal and stress analysis, it is evident that the ceramic LLL is a crucial weak point of the W DLP. It not only limits the melting time threshold but also causes crack problems during EAST 600kA discharge events. All the analysis results show a fact that lower $I_p$ is good for W DLP's life. Our calculation results can not only give important advice for EAST present discharge operation but also provide useful reference for the future improvement of the W DLP.